\documentclass[a4paper,conference]{IEEEtran}
\usepackage{amsmath,amssymb,amsfonts}
\usepackage{algorithmic,microtype}
\usepackage{graphicx}
\usepackage{dsfont}
\usepackage{textcomp}
\usepackage{mathtools}
\usepackage{array}
\usepackage{algorithmic}
\usepackage{algorithm,color}
\usepackage{tabularx}
\usepackage{multirow}
\newcolumntype{L}[1]{>{\raggedright\let\newline\\\arraybackslash\hspace{0pt}}m{#1}}
\newcolumntype{C}[1]{>{\centering\let\newline\\\arraybackslash\hspace{0pt}}m{#1}}
\newcolumntype{R}[1]{>{\raggedleft\let\newline\\\arraybackslash\hspace{0pt}}m{#1}}

\def\BibTeX{{\rm B\kern-.05em{\sc i\kern-.025em b}\kern-.08em
    T\kern-.1667em\lower.7ex\hbox{E}\kern-.125emX}}

\begin{document}
\linespread{1}
\title{\LARGE Study of MMSE-Based Resource Allocation for Clustered Cell-Free Massive MIMO Networks}

\author{Saeed Mashdour$^{\star}$ and Rodrigo C. de Lamare $^{\star,\dagger}$,  $^{\ddagger}$ \\  $^{\star}$ Centre for Telecommunications Studies, Pontifical Catholic University of Rio de Janeiro, Brazil \\
$^{\dagger}$ Department of Electronic Engineering, University of York, United Kingdom \\
smashdour@gmail.com, delamare@cetuc.puc-rio.br  \vspace{-2em}\\
\thanks{This work was supported by CNPQ, CPQD and FAPERJ.}}

\maketitle

\begin{abstract}
In this paper, a downlink cell-free massive multiple-input multiple-output (CF massive MIMO) system and a network clustering is considered. Closed form sum-rate expressions are derived for CF and the clustered CF (CLCF) networks where linear precoders included zero forcing (ZF) and minimum mean square error (MMSE) are implemented. An MMSE-based resource allocation technique with multiuser scheduling based on an enhanced greedy technique and power allocation based on the gradient descent (GD) method is proposed in the CLCF network to improve the system performance. Numerical results show that the proposed technique is superior to the existing approaches and the computational cost and the signaling load are essentially reduced in the CLCF network.\\
\end{abstract}

\begin{IEEEkeywords}
Massive MIMO, cell-free, clustering, multiuser scheduling, power allocation. \vspace{-1em}
\end{IEEEkeywords}

\footnote{This work was supported by Funttel/Finep - Grant No. 01.20.0179.00. The authors also thank the support of CNPq, FAPESP and FAPERJ.}

\section{Introduction}
Cell-free (CF) massive multiple-input multiple-output (MIMO) networks were initially proposed in \cite{ngo2015cell, ngo2017cell} where a
large number of access points (APs) jointly
serve all user equipments (UEs) using the same time frequency resources through a wide area. Thereby, a  higher coverage rate and throughput could be provided compared to cellular networks. Regarding the huge burden imposed on the processing unit in CF networks, network clustering including user-centric and network-centric approaches are suggested in literature to make it practical in terms of signaling and computational complexity  \cite{bjornson2020scalable, demir2021foundations, tentu2022uav, guevara2021partial, wei2022user}.

The downlink of MIMO networks \cite{mmimo,wence} require transmit processing techniques such as precoding \cite{siprec,gbd,zfsr,rmmse,locsme,oskpme,baplnc,1bitprec,wlbd,mbthp,rmbthp,cqabd,cfrmmse,rsrbd,rsthp,zcprec} and resource allocation \cite{tdscl,tds,jpba,lrcc,cqabd,rapa}. In particular,  one of the most important concepts in CF networks in practice is resource allocation including multiuser scheduling and power allocation to improve  network performance. Multiuser scheduling is fundamental in multiuser interference reduction \cite{interdonato2019ubiquitous} and power allocation is crucial to gain the desirable performance. These challenges are addressed in several works. In \cite{mosleh2019downlink}, downlink resource allocation in CF massive MIMO networks is studied through the maximization of the minimum achievable rate among the UEs. The work of \cite{wu2021revenue} has studied a network slicing based CF massive MIMO architecture to support many UEs through resource slicing in a virtualized wireless CF massive MIMO network. A weighted
sum-rate (WSR) problem was solved in \cite{ammar2021downlink} using fractional programming and employing compressive sensing for multiuser scheduling and power allocation in clustered CF (CLCF) massive MIMO networks.

In this paper, we investigate the downlink of CF and CLCF massive MIMO networks and investigate resource allocation techniques for sum-rate maximization. In particular, we derive closed form expressions for sum-rates of CF and CLCF massive MIMO networks and develop a sequential multiuser scheduling and power allocation (SMSPA) scheme. For the SMSPA scheme, we develop an improved greedy subset selection (IGSS) technique based on the algorithm proposed in \cite{mashdour2022multiuser} for user scheduling and devise a gradient descent (GD) power allocation algorithm for mean square error (MSE) minimization which is equivalent to sum-rate maximization as will be described later. Simulation results show the superiority of the proposed SMSPA scheme and resource allocation techniques as compared to existing methods.

{\it Notation}: Throughout the paper, $\left \| . \right \|_{F}$ denotes the Frobenius norm, $\textbf{I}_{N}\in \mathbb{C}^{N\times N}$ denotes the identity matrix, the complex normal distribution is represented by $\mathcal{CN}\left ( .,. \right )$, superscripts $^{T}$, $^{\ast}$, and $^{H}$ denote transpose,
complex conjugate and hermitian operations respectively, $\mathcal{A}\cup \mathcal{B} $ is union of sets $\mathcal{A}$ and $\mathcal{B}$, $\mathcal{A}\setminus \mathcal{B} $ shows exclusion of set $\mathcal{B}$ from set $\mathcal{A}$, and ${Tr}\left ( . \right )$ shows the Trace operator.

\section{System Model}
We consider a CF network including $M$ randomly distributed single antenna APs and $K$ uniformly distributed single antenna UEs. The CLCF network is formed by dividing the CF network into $C$ non-overlapping areas where the cluster $c$ includes $M_c$ randomly distributed single antenna APs and $K_{c}$ uniformly distributed single antenna UEs. We assume that the number of UEs is much larger than the number of APs in both networks so that $K> > M$.

\subsection{CF and CLCF Networks}\label{AA}
In the CF network including $M$ APs and and $K$ UEs, the channel between AP $m$ and UE $k$ is shown by the coefficient $g_{mk}=\sqrt{\beta _{mk}}h_{mk}$ \cite{ngo2017cell}, where $\beta _{mk}$ are the large-scale fading coefficients due to the path loss and shadowing and $ h_{mk}\sim \mathcal{CN}\left ( 0,1 \right )$ denote the independent and identically distributed (i.i.d.) random variables (RVs) modelling small-scale fading
that remain constant during a coherence interval and which are assumed to be independent over different coherence intervals. Further, we assume the following model of the large scale coefficients  $\beta _{mk}=\textup{PL}_{mk} \times 10^{\frac{\sigma _{sh}z_{mk}}{10}}$ where $\textup{PL}_{mk}$ is the path loss and $10^{\frac{\sigma _{sh}z_{mk}}{10}}$
refers to  shadow fading where $\sigma _{sh}=8\textup{dB}$,  $z_{mk}\sim\mathcal{N}\left ( 0,1 \right )$. Considering $d_{mk}$ is the distance between the AP $m$ and UE $k$, the path loss is modeled using the results from \cite{tang2001mobile} as
\begin{equation}
    \textup{PL}_{mk}=\left\{\begin{matrix}
-\textup{D}-35\log_{10}\left ( d_{mk} \right ), \textup{if }  d_{mk}>d_{1}& \\
 -\textup{D}-10\log_{10}\left ( d_{1}^{1.5} d_{mk}^2\right ), \textup{if }  d_{0}<d_{mk}\leq d_{1}& \\
 -\textup{D}-10\log_{10}\left ( d_{1}^{1.5} d_{0}^2 \right ), \textup{if }  d_{mk}\leq d_{0}&
\end{matrix}\right.,
\end{equation}
where
\begin{equation}
\begin{multlined}
    \textup{D}=46.3+33.9\log_{10}\left ( f \right )-13.82\log_{10}\left ( h_{AP} \right ) \\
 -\left [ 1.11\log_{10}\left ( f \right )-0.7 \right ]h_{u}+1.56\log_{10}\left ( f \right )-0.8
\end{multlined}
\end{equation}
$f=1900$MHz is the carrier frequency, $h_{AP}=$15m and $h_{u}=$1.5m are the
AP and UE antenna heights, respectively, $d_{0}=$10m and $d_{1}=$50m. When $d_{mk}\leq d_{1},$ there is no shadowing. The downlink signal received at $n$ scheduled UEs is given by
\begin{equation}
    \mathbf{y}=\sqrt{\rho _{f}}\mathbf{G}^T\mathbf{P}\mathbf{x}+\mathbf{w}
\end{equation}
where $\rho _{f}$ is the maximum transmitted power of each antenna, $\mathbf{G}=\hat{\textbf{G}}+\tilde{\textbf{G}}\in \mathbb{C}^{M\times n}$ is the channel matrix with the channel estimate $\hat{\textbf{G}}$ and the estimation error $\tilde{\textbf{G}}$ which models the CSI imperfection, $\left [ \mathbf{G} \right ]_{m,k}=g_{mk}$, $\mathbf{P}\in \mathbb{C}^{M\times n}$ is the linear precoder matrix, $\mathbf{x}=\left [ x_{1},\cdots ,x_{n} \right ]^{T}$ is the zero mean symbol vector, $\mathbf{x}\sim \mathcal{CN}\left ( \mathbf{0},\mathbf{I}_{n} \right )$, and $\mathbf{w}=\left [ w_{1},\cdots,w_{n}  \right ]^{T}$ is the additive noise vector, $\textbf{w}\sim \mathcal{CN}\left ( 0,\sigma_{w}^{2}\textbf{I}_{n} \right )$. We consider elements of $\mathbf{x}$ to be mutually independent, and independent of all noise and channel coefficients. Thus, we can obtain the sum-rate of the CF system as follows
\begin{equation}\label{eq:RCF}
    SR_{cf}=\log_{2}\left (  \det\left [\textbf{R}_{cf}+\textbf{I}_n  \right ]\right )
\end{equation}
where the covariance matrix $\textbf{R}_{cf}$ is given by
\begin{equation}\label{eq:RCF_1}
    \textbf{R}_{cf}=\rho _{f} \hat{\textbf{G}}^{T}\textbf{P}\textbf{P}^{H}\hat{\textbf{G}}^{\ast }\left ( \rho _{f}\tilde{\textbf{G}}^{T}\textbf{P}\textbf{P}^{H}\tilde{\textbf{G}}^{\ast } +\sigma _{w}^{2}\textbf{I}_n\right )^{-1}
\end{equation}
When clustering is considered, the downlink received signal at $n_{c}$ scheduled UEs of the cluster $c$ is given by
\begin{equation} \label{yc}
\textbf{y}_{c}=\sqrt{\rho _{f}}{\textbf{G}}_{cc}^T\textbf{P}_{c}\textbf{x}_{c}+
\sum_{i=1,i\neq c}^{C}\sqrt{\rho _{f}}{\textbf{G}}_{ic}^T\textbf{P}_{i}\textbf{x}_{i}+\textbf{w}_{c}
\end{equation}
where $\textbf{G}_{ic}=\hat{\textbf{G}}_{ic}+\tilde{\textbf{G}}_{ic}\in \mathbb{C}^{M_i\times n_{c}}$ is the channel from APs of the cluster $i$ to the UEs of the cluster $c$, $\textbf{P}_{i}\in \mathbb{C}^{M_i\times n_{c}}$ is the linear precoding matrix, and  $\textbf{x}_i=\left [ x_{i1},\cdots ,x_{in_{c}} \right ]^{T}$ is the symbol vector of the cluster $i$, $\textbf{x}_i\sim \mathcal{CN}\left ( \mathbf{0},\mathbf{I}_{n_{c}} \right )$, $i\in \left \{ 1, 2, \cdots ,C \right \}$, and $\textbf{w}_c= [w_{c_{1}}, \cdots, w_{c_{n_{c}}} ]^{T}$ is the additive noise vector, $\textbf{w}_c\sim \mathcal{CN}\left ( 0,\sigma_{w}^{2}\textbf{I}_{n_{c}} \right )$. Accordingly, the sum-rate expression for the cluster $c$ is obtained as
\begin{equation}\label{eq:RCL0}
    SR_{c}=\log_{2}\left (  \det\left [\left ( \rho _{f}\hat{\textbf{G}}_{cc}^T\textbf{P}_{c}\textbf{P}_{c}^{H}\hat{\textbf{G}}_{cc}^* \right )\textbf{R}_{c}^{-1}+\textbf{I}_{n_{c}}  \right ]\right )
\end{equation}
and the covariance matrix $\textbf{R}_{c}$ is described by
\begin{equation}\label{eq:RCL1}
\begin{split}
\textbf{R}_{c}
&=\rho _{f}\tilde{\textbf{G}}_{cc}^T\textbf{P}_{c}\textbf{P}_{c}^{H}\tilde{\textbf{G}}_{cc}^*+\sum_{i=1,i\neq c}^{C}\rho _{f}\hat{\textbf{G}}_{ic}^T\textbf{P}_{i}\textbf{P}_{i}^{H}\hat{\textbf{G}}_{ic}^*+\\
&\sum_{i=1,i\neq c}^{C}\rho _{f}\tilde{\textbf{G}}_{ic}^T\textbf{P}_{i}\textbf{P}_{i}^{H}\tilde{\textbf{G}}_{ic}^*+\sigma _{w}^{2}\textbf{I}_{n_{c}}
\end{split}
\end{equation}
where $\textbf{x}_c$ and $\textbf{w}_c$ are assumed to be statistically independent. Thus, the total network sum-rate is achieved by summation of Equation (\ref{eq:RCL0}) over all  clusters. At the receiver of the UEs several detection techniques \cite{jidf,spa,mfsic,mbdf,bfidd,cpm,dynovs} can be used.

\section{Proposed Resource Allocation Technique}

For resource allocation in the CLCF network, in the proposed scheme, we first accomplish the multiuser scheduling according to the method implemented in \cite{mashdour2022enhanced}, which determines various UE sets based on an enhanced greedy technique presented in \cite{mashdour2022multiuser} and considers equal power loading in the assignment of the sets and finally selects the best set according to the sum-rate criterion. Thereafter, a power allocation algorithm is proposed based on the gradient descent algorithm for MSE minimization. The proposed power allocation method significantly improves the sum-rate performance because MSE minimization is
equivalent to the maximization of the Signal to interference plus noise ratio (SINR) which in turn is equivalent
to the sum-rate maximization when Gaussian signaling is
used \cite{verdu1998multiuser}. Note that the proposed technique could be extended to be used in  CF networks without clustering as well.

\subsection{User Scheduling Algorithm }\label{User Scheduling}
Since the number of UEs is much larger than the number of APs, $K_c>>M_c$, we select the first set of UEs by adapting the implemented greedy method in \cite{dimic2005downlink} so that we can schedule a set of UEs with a pre-determined length. We consider an MMSE precoder unlike the ZF precoder used in \cite{dimic2005downlink} to improve the performance and the search algorithm. Considering $n_{c}$ as length of the selected set $\mathcal{S}_{n_{c}}$ and consequently the channel matrix $\textbf{G}_{cc}\left ( \mathcal{S}_{n_{c}} \right )\in \mathbb{C}^{M_{c}\times {n_{c}}}$, the algorithm is the solution to the following problem
\begin{equation}
\begin{aligned}
& {\max_{\mathcal{S}_{n_{c}} } ~SR_{MMSE} ( {\mathcal{S}}_{n_{c}}}  ) \\
& \text{subject to} ||\textbf{P}_c ( \mathcal{S}_{n_{c}} ) ||_{F}^{2}\leq P.
\end{aligned}
\end{equation}
where $SR_{MMSE}( {\mathcal{S}}_{n_{c}})$ is the sum-rate when MMSE precoder is used, $P$ is the upper limit of the signal covariance matrix ${Tr}[ \textbf{C}_{\textbf{x}}]\leq P$ and $\textbf{P}_c (\mathcal{S}_{n_{c}}  ) \in \mathbb{C}^{M_{c}\times {n_{c}}}$ is the precoding matrix. In the case of power allocation, the precoding matrix is defined as $\textbf{P}_c ( \mathcal{S}_{n_{c}}) = \textbf{W}_{c}\textbf{D}_{c}$ where $\textbf{W}_{c}\in\mathbb{C}^{M_c \times {n_{c}}}$ is the {normalized MMSE weight matrix} and $\textbf{D}_{c}\in\mathbb{C}^{{n_{c}} \times {n_{c}}}$ is the power {allocation matrix} considered as
\begin{multline}
\textbf{D}_{c}=\begin{bmatrix}
\sqrt{p_{1}} & 0 & \cdots  & 0\\
 0& \sqrt{p_{2}} & \cdots &0 \\
 \vdots & \vdots  &\cdots   & \vdots \\
 0&0  &\cdots   & \sqrt{p_{{n_{c}}} }
\end{bmatrix}=\textup{diag}\left ( \textbf{d}_c \right )\\
, \textbf{d}_c
=\left [ \sqrt{p_{1}} \ \sqrt{p_{2}} \ \cdots \  \sqrt{p_{{n_{c}}}} \right ]^T
.\end{multline}
In order to evaluate more sets of UEs so that we can approach the optimal set chosen by  exhaustive search, we remove one UE from the set $\mathcal{S}_{{n_{c}}\left ( 1 \right )}$ as the UE with the lowest channel power among the selected set and substitute it with another UE with the highest channel power from the remaining UEs other than $\mathcal{S}_{{n_{c}}\left ( 1 \right )}$ to achieve the second set as $\mathcal{S}_{{n_{c}}\left ( 2 \right )}$. The removed and the substituting UEs are considered as follows, respectively,
\begin{equation} \label{kex}
         k_{r\left (1\right )}=\underset{k\in \mathcal{S}_{{n_{c}}\left ( 1 \right )}}{{\arg \min}}~ \textbf{g}_{k}^{H}\textbf{g}_{k}
\end{equation}
\begin{equation} \label{knew}
 k_{su\left (1\right )}=\underset{k\in \mathcal{K}_{re\left ( 1\right )}}{{\arg \max}}~\textbf{g}_{k}^{H}\textbf{g}_{k}
\end{equation}
where $\textbf{g}_{k}\in \mathbb{C}^{M_c \times 1}$ is the channel vector to UE $k$ and $\mathcal{K}_{re\left ( 1\right )}=\mathcal{K}_c\setminus \mathcal{S}_{{n_{c}}\left ( 1 \right )}$is the set of remaining UEs other than the set $\mathcal{S}_{{n_{c}}\left ( 1 \right )}$, $\mathcal{K}_c=\left \{ 1,2,\cdots ,K_c \right \}$. Doing the same procedure for $\mathcal{S}_{{n_{c}}\left ( 2 \right )}$, we obtain $\mathcal{S}_{{n_{c}}\left ( 3 \right )}$ and so on until we achieve $K_c-n_{c}+1$ sets. For ${i \in \left \{ 2,\ldots ,{K_{c}-n_{c}}+1 \right \}}$, the UE set and the set of remaining UEs  are shown as follows, respectively,
\begin{equation}
    \mathcal{S}_{{n_{c}}\left ( i \right )}=\left (\mathcal{S}_{{n_{c}}\left ( i-1 \right )}\setminus k_{r\left (i-1\right )} \right )\cup k_{su\left (i-1\right )},
\end{equation}
\vspace{-4mm}
\begin{equation}
    \mathcal{K}_{re\left ( i\right )}=\mathcal{K}_{re\left ( i-1\right )}\setminus k_{su\left (i-1\right )}.
\end{equation}
Finally, the desired scheduled set of UEs $\mathcal{S}_{{n_{c}}_{d}}$ is the set which results in the highest sum-rate among the UE sets $\mathcal{S}_{n_c\left ( m\right )}, {m \in \left \{ 1,\ldots ,{K_{c}-{n_{c}}}+1 \right \}}$. Algorithm \ref{alg:alg1} explains the proposed user scheduling approach in detail.

\begin{algorithm}[H]
\caption{Proposed IGSS User Scheduling Algorithm.}\label{alg:alg1}
\begin{algorithmic}
\STATE
\vspace{-2mm}
\STATE {{j=1}} \ \% selection of the first set
\STATE \hspace{0.5cm} \textbf{set} $l = 1$
\STATE \hspace{0.5cm} \textbf{find a user such that}
\STATE \hspace{0.5cm} $k_{1}=\underset{k\in \mathcal{K}_c}{\textup{argmax}} ~\textbf{g}_{k}^{H}\textbf{g}_{k}$
\STATE \hspace{0.5cm} \textbf{set $\mathcal{U}_1 = {k_1}$ and denote the achieved rate}
\STATE \hspace{0.5cm} $SR_{MMSE}\left ( \mathcal{U}_{1} \right )$
\STATE \hspace{0.5cm} \textbf{while 
{$l<{n_{c}}$}}
\STATE \hspace{1cm} $l=l+1$
\STATE \hspace{1cm} \textbf{find a user $k_{l}$ such that}
\STATE \hspace{1cm} $k_{l}=\underset{k\in\left (  \mathcal{K}_c \setminus  \mathcal{U}_{l-1} \right )}{\textup{argmax}}SR_{MMSE}\left ( \mathcal{U}_{l-1}\cup \left \{ k \right \} \right )$
\STATE \hspace{1cm} \textbf{set $\mathcal{U}_{l}=\mathcal{U}_{l-1}\cup \left \{ k_{l} \right \}$ and denote the rate}
\STATE \hspace{1cm} $SR_{MMSE}\left ( \mathcal{U}_{l} \right )$
\STATE \hspace{1cm} \textbf{If $SR_{MMSE}\left ( \mathcal{U}_{l} \right )\leq SR_{MMSE}\left ( \mathcal{U}_{l-1} \right )$, break}
\STATE \hspace{1cm}  $l=l-1$
\STATE \hspace{0.5cm} $\mathcal{S}_{{n_{c}}\left ( \textup{j} \right )}=\mathcal{U}_{l}$
\STATE \hspace{0.5cm} \textbf{compute}: $ SR_{c}\left ( \mathcal{S}_{{n_{c}}\left ( \textup{j} \right )}\right )$
\STATE \hspace{0.5cm} $\mathcal{K}_{re\left ( \textup{j}\right )}=\mathcal{K}_c\setminus \mathcal{S}_{{n_{c}}\left ( \textup{j} \right )}$ \
\STATE \hspace{0.5cm} $ k_{r\left (\textup{j}\right )}=\underset{k\in \mathcal{S}_{{n_{c}}\left ( \textup{j} \right )}}{\textup{argmin}}\textbf{g}_{k}^{H}\textbf{g}_{k}$
\STATE \hspace{0.5cm} $k_{su\left (\textup{j}\right )}=\underset{k\in \mathcal{K}_{re\left ( \textup{j}\right )}}{\textup{argmax}}\textbf{g}_{k}^{H}\textbf{g}_{k}$
\STATE \textbf{for} $\textup{j}=2$ \textbf{to} 
{${K_{c}-{n_{c}}}+1$}
\STATE \hspace{0.5cm} $\mathcal{S}_{{n_{c}}\left ( \textup{j} \right )}=\left (\mathcal{S}_{{n_{c}}\left ( \textup{j}-1 \right )}\setminus k_{r\left (\textup{j}-1\right )}  \right )\cup k_{su\left (\textup{j}-1\right )}$
\STATE \hspace{0.5cm} $\mathcal{K}_{re\left ( \textup{j}\right )}=\mathcal{K}_{re\left ( \textup{j}-1\right )}\setminus k_{su\left (\textup{j}-1\right )}$
\STATE \hspace{0.5cm} $ k_{r\left (\textup{j}\right )}=\underset{k\in \mathcal{S}_{{n_{c}}\left ( \textup{j} \right )}}{\textup{argmin}}\textbf{g}_{k}^{H}\textbf{g}_{k}$
\STATE \hspace{0.5cm} $k_{su\left (\textup{j}\right )}=\underset{k\in \mathcal{K}_{re\left ( \textup{j}\right )}}{\textup{argmax}}\textbf{g}_{k}^{H}\textbf{g}_{k}$
\STATE \hspace{0.5cm} \textbf{compute}: $SR_{c}\left ( \mathcal{S}_{{n_{c}}\left ( \textup{j} \right )} \right )$
\STATE \textbf{end for}
\STATE $\mathcal{S}_{{n_{c}}_{d}}=\underset{\mathcal{S}_{{n_{c}}} \in \mathcal{S}_{{n_{c}}\left ( {m} \right )}}{\textup{argmax}}\left \{  SR_{c}\left ( \mathcal{S}_{{n_{c}}} \right ) \right \}$
\STATE \textbf{Precoding of $\mathcal{S}_{{n_{c}}_{d}}$}
\end{algorithmic}
\label{alg1}
\end{algorithm}
\subsection{Gradient Descent Power Allocation Algorithm}\label{Gradient Descent}
Since the estimated received signal of the cluster $c$ of Equation (\ref{yc}) consists of a desired term and the terms associated with imperfect CSI, inter-cluster interference and  noise, applying power allocation, we rewrite the estimated signal as follows
\begin{equation} \label{yc2}
\begin{split}
\textbf{y}_{c}&=\sqrt{\rho _{f}}\hat{\textbf{G}}_{cc}^T\textbf{W}_{c}\textbf{D}_{c}\textbf{x}_{c}+\sqrt{\rho _{f}}\tilde{\textbf{G}}_{cc}^T\textbf{P}_{c}\textbf{x}_{c}+\\
&\sum_{i=1,i\neq c}^{C}\sqrt{\rho _{f}}\textbf{G}_{ic}^T\textbf{P}_{i}\textbf{x}_{i}+\textbf{w}_{c}
\end{split}
\end{equation}
We use the MSE between the transmitted signal and the estimated signal at the receiver as the objective function because MSE minimization is equivalent to sum-rate maximization as described before. Therefore, the following power allocation problem is defined:
\begin{equation} \label{opt.1}
\begin{aligned}
& \underset{\textbf{d}_{c}}{\text{min}}~\mathbb{E}\left [ \varepsilon  \right ] \\
& \text{subject to}\ \left \| \textbf{W}_c \textup{diag}\left ( \textbf{d}_{c} \right ) \right \|^{2}\leq P
\end{aligned}
\end{equation}
where the error is
\begin{equation}
    \varepsilon =\left \| \textbf{x}_{c}-\textbf{y}_{c} \right \|^{2}
\end{equation}
and
\begin{equation}
    \begin{split}
    \textbf{x}_{c}-\textbf{y}_{c}&=\textbf{x}_{c}-\sqrt{\rho _{f}}\hat{\textbf{G}}_{cc}^T\textbf{W}_{c}\textup{diag}\left ( \textbf{d}_{c}\right )\textbf{x}_{c}\\
    &-\sqrt{\rho _{f}}{\tilde{\textbf{G}}}_{cc}^T\textbf{P}_{c}\textbf{x}_{c}
    -\sum_{i=1,i\neq c}^{C}\sqrt{\rho _{f}}\textbf{G}_{ic}^T\textbf{P}_{i}\textbf{x}_{i}-\textbf{w}_{c}
\end{split}
\end{equation}
Thus, we can evaluate the error as in (\ref{opt.1}). Since this
error is scalar, it remains the same when the trace operator is
applied over the right hand side. Then, using the property
${Tr}\left ( \textbf{A}+\textbf{B} \right )={Tr}\left ( \textbf{A}
\right )+{Tr}\left ( \textbf{B} \right )$, where $\textbf{A}$ and
$\textbf{B}$ are two equal dimension matrices, the error is
rewritten different ways that can be more convenient for
manipulation.

Considering the transmitted signal of each cluster uncorrelated with
the transmitted signal from other clusters, the expected value of
the error in (\ref{opt.1}) is expressed in other forms. We then take
the derivative of the error with respect to the power loading matrix
$\textbf{D}_{c}$ and the equality $\frac{\partial {Tr}\left (
\textbf{AB} \right )}{\partial \textbf{A}}=\textbf{B}\odot
\textbf{I}$ is used where $\textbf{A}$ is a diagonal matrix and
$\odot$ shows the Hadamard product. Thus, we obtain
 \begin{equation} \label{first.or}
 \begin{split}
     \frac{\partial \mathbb{E}\left ( \varepsilon  \right ) }{\partial \textbf{D}_{c}}&=2\rho _{f}\left ( \textbf{W}_{c}^{H}\hat{\textbf{G}}_{cc}^*\hat{\textbf{G}}_{cc}^T\textbf{W}_{c}\textup{diag}\left ( \textbf{d}_{c}\right ) \right )\odot \textbf{I}-\\
     &\sqrt{\rho _{f}}\hat{\textbf{G}}_{cc}^T\textbf{W}_{c}\odot \textbf{I}-\sqrt{\rho _{f}}\textbf{W}_{c}^{H}\hat{\textbf{G}}_{cc}^*\odot \textbf{I}+\\
     &\rho_{f}\textbf{W}_{c}^{H}\hat{\textbf{G}}_{cc}^*{\tilde{\textbf{G}}}_{cc}^T\textbf{P}_{c}\odot \textbf{I}+\rho_{f}\textbf{P}_{c}^{H}\tilde{\textbf{G}}_{cc}^*\hat{\textbf{G}}_{cc}^T\textbf{W}_{c}\odot \textbf{I}=\\
     &2\rho _{f}\left ( \textbf{W}_{c}^{H}\hat{\textbf{G}}_{cc}^*\hat{\textbf{G}}_{cc}^T\textbf{W}_{c}\textup{diag}\left ( \textbf{d}_{c}\right ) \right )\odot \textbf{I}-\\
     &2\sqrt{\rho _{f}}Re\left \{ \textbf{W}_{c}^{H}\hat{\textbf{G}}_{cc}^*\odot \textbf{I} \right \}\\
     &+2{\rho _{f}}Re\left \{ \textbf{W}_{c}^{H}\hat{\textbf{G}}_{cc}^*{\tilde{\textbf{G}}}_{cc}^T\textbf{P}_{c}\odot \textbf{I} \right \}
     \end{split}
 \end{equation}
 where $Re\left \{ \textbf{A} \right \}$ shows the real part of matrix $\textbf{A}$. We can use a stochastic gradient descent approach to update the power allocation coefficient as follows:
 \begin{equation}
 \begin{split}
     \textbf{d}_{c}\left ( i \right )&=\textbf{d}_{c}\left ( i-1 \right )-\lambda \frac{\partial \mathbb{E}\left ( \varepsilon  \right ) }{\partial \textbf{D}_{c}}=\\
     &\textbf{d}_{c}\left ( i-1\right )-\\
     &2\rho _{f} \lambda \left ( \textbf{W}_{c}^{H}\hat{\textbf{G}}_{cc}^*\hat{\textbf{G}}_{cc}^T\textbf{W}_{c}\textup{diag}\left ( \textbf{d}_{c}\left ( i-1\right )\right ) \right )\odot \textbf{I}+\\
     &2\sqrt{\rho _{f}} \lambda Re\left \{ \textbf{W}_{c}^{H}\hat{\textbf{G}}_{cc}^*\odot \textbf{I} \right \}-\\
     &2{\rho _{f}}\lambda Re\left \{ \textbf{W}_{c}^{H}\hat{\textbf{G}}_{cc}^*{\tilde{\textbf{G}}}_{cc}^T\textbf{P}_{c}\odot \textbf{I} \right \}
     \end{split}
 \end{equation}
 where $i$ is the iteration index, and $\lambda$ is the positive step size. 
 {Before running the adaptive algorithm, the transmit power constraint should be satisfied so that $\left \| \textbf{W}_c \textup{diag}\left ( \textbf{d}_{c} \right ) \right \|^{2}=\left \| \textbf{P}_{c} \right \|^{2}\leq P$ where $\textbf{P}_{c}$ is the precoding matrix, $\textbf{W}_c$ is the normalized precoding matrix and $\textup{diag}\left ( \textbf{d}_{c} \right )$ is the power allocation matrix.} Therefore, the power scaling factor 
 {$\eta =\sqrt{\frac{{Tr}\left ( \textbf{P}_{c}\textbf{P}_{c}^{H} \right )}{{Tr}\left ( \textbf{W}_{c}\textup{diag}\left ( \textbf{d}_{c}.\textbf{d}_{c} \right )\textbf{W}_{c}^{H} \right )}}$}  is employed in each iteration to scale the coefficients properly. 

\section{Simulation Results}

In this section, we assess the sum-rate performances of the CF and CLCF scenarios for the proposed SMSPA scheme and algorithms and the existing techniques including exhaustive search (ExS), greedy (Gr) and the WSR method proposed in \cite{ammar2021downlink} which is adapted to the proposed clustering by maximizing WSR for the UEs of the clusters supported by the corresponding APs. The CF network is particularly considered as a squared area with the side length of 400m including $M$ single-antennas randomly located APs and $K$ uniformly distributed single antenna UEs. For a CLCF network, we consider the same area divided into $C=4$ non-overlapping clusters so that cluster $c$ includes $M_c$ single-antennas randomly located APs and $K_{c}$ uniformly distributed single antenna UEs. To summarize, the main simulation parameters are described in Table~\ref{table:simu-parameters}.

\begin{table}[htb!]
\begin{small}
\caption{Simulation parameters}
\begin{center}
\begin{tabular}{| l | l |}
\hline
Parameter & Value\\ [0.5ex]
\hline
\hline
Carrier frequency & 1900 MHz \\
\hline
AP antenna height & 15 m \\
\hline
UE antenna height & 1.5 m \\
\hline
Shadowing factor & 8 dB \\
\hline
Area size & 400 m$^{2}$ \\
\hline
NO of clusters & 4 \\
\hline
Symbol energy & $P_s$=1 \\
\hline
Transmit power & $M\times P_{s}$ \\
\hline
\end{tabular}
\label{table:simu-parameters}
\end{center}
\end{small}
\end{table}

\begin{figure}
    \centering
        \includegraphics[width=.8\linewidth]{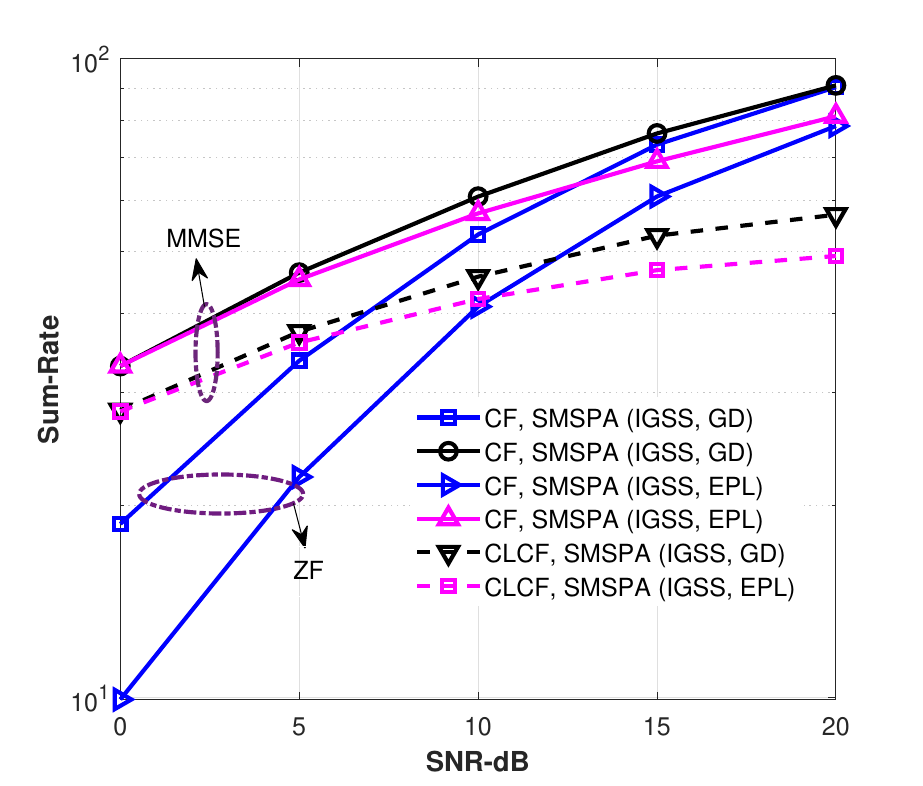}
        \vspace{-0.75em}
    \caption{\small{Comparison of the proposed SMSPA technique in CF and CLCF networks with the system that uses the proposed IGSS scheme and EPL for ZF and MMSE precoders, ($M=64$, $K=128$, $n=24$)}}
    \label{fig:fig1}
\end{figure}

 To highlight the power allocation effect, the performance of the proposed resource allocation technique is compared with the system which has employed the proposed user scheduling and equal power loading (EPL) for CF and CLCF networks in Fig. \ref{fig:fig1} while ZF and MMSE precoders are used. As it is visible, sum-rates improve by increasing the signal to noise ratio (SNR) and MMSE precoder has outperformed ZF. We can also note that implementing the proposed SMSPA method has significantly improved the performance against the case of equal power loading (EPL) in both CF and CLCF networks. In addition, the CF network shows a better performance compared to CLCF which is in result of the additional interference caused by other clusters as shown in Equation (\ref{eq:RCL1}). This is while the CLCF network substantially reduces the computational complexity and the signaling load compared to CF network as shown in Table \ref{table:complexity}.

\begin{figure}
    \centering
        \includegraphics[width=.8\linewidth]{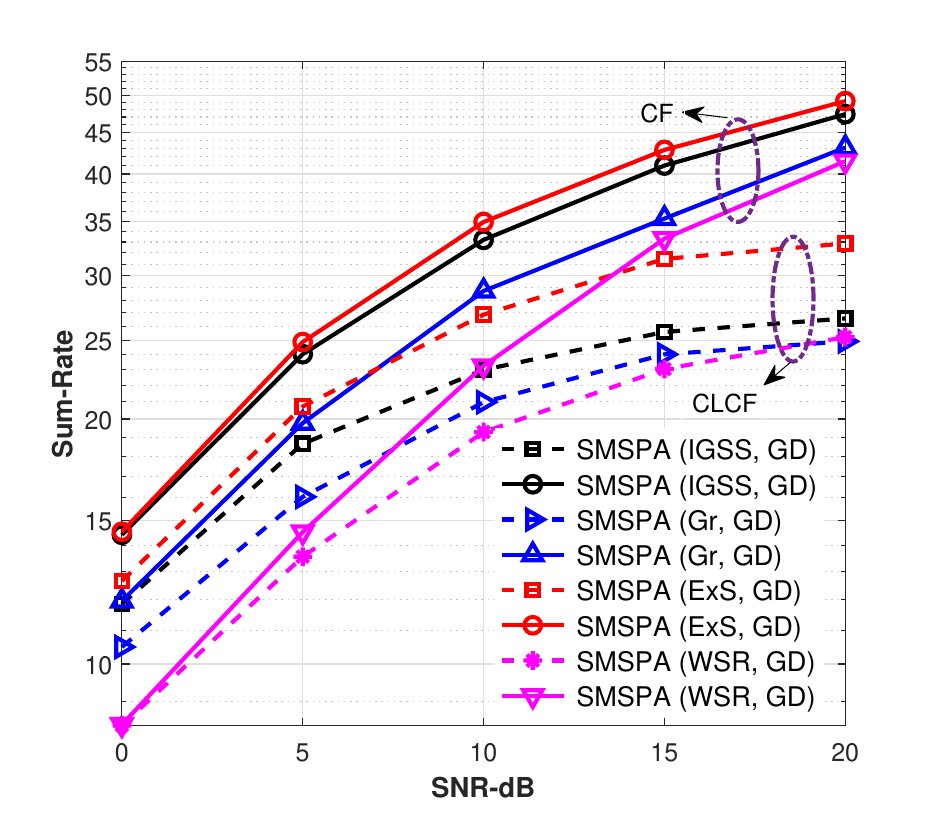}
        \vspace{-0.75em}
    \caption{\small{Comparison of different resource allocation techniques considering a GD power allocation algorithm in CF and CLCF networks when an MMSE precoder is used, ($M=64$, $K=16$, $n=8$)}}
    \label{fig:fig2}
\end{figure}

 In Fig.~\ref{fig:fig2}, using MMSE precoder, we have compared the proposed SMSPA resource allocation technique in CF and CLCF networks with different techniques where the proposed GD algorithm is implemented for power allocation. However, in order to consider the comparison with the optimal ExS method, we have examined a small number of UEs in the network and half of the UEs are scheduled. We can note that the proposed SMSPA algorithm has outperformed other techniques and the results approach the optimal ExS method.


\begin{table}[htb!]
\begin{small}
\caption{Computational complexity of the proposed resource allocation algorithm and the signaling load for CF and CLCF networks}
\begin{center}
\begin{tabular}{| m{3em} | m{6em} | m{4em} | m{6em} |}
\hline
Network &NO of APs, UEs and scheduled UEs &Signaling load & NO of FLOPs\\ [0.5ex]
 \hline
\hline
CF& $M=64$, $K=16$ and $n=8$& 3072 &$0.78 \times 10^{6}$\\
\hline
CLCF& $M=64$, $K=16$ and $n=8$& 768 &$0.75 \times 10^{5}$\\
\hline
CF& $M=128$, $K=256$ and $n=128$& 98304 &$2.1643 \times 10^{10}$\\
\hline
CLCF & $M=128$, $K=256$ and $n=128$& 24576 &$1.1073 \times 10^{9}$\\
\hline
\end{tabular}
\label{table:complexity}
\end{center}
\end{small}
\end{table}

\section{Conclusion}

This paper has developed an SMSPA resource allocation technique for
downlink CLCF massive MIMO networks including IGSS multiuser
scheduling algorithm which adapts an enhanced greedy technique and a
power allocation algorithm which uses a GD method. Simulation
results show that the MMSE precoder outperforms ZF and the GD power
allocation significantly enhances the sum-rates. It is also shown
that the proposed SMSPA technique has better performance than
existing methods to a large extent so that the sum-rate performance
approaches the optimal exhaustive search. Moreover, the network
clustering significantly saves cost and complexity.

\bibliographystyle{IEEEbib}
\bibliography{refs}

\end{document}